\documentclass[12pt]{iopart}
\usepackage{psfig}
\usepackage{graphicx}
\begin{document}

\article[Fast motions in Coma I cloud]{}{Fast motions of galaxies in the Coma I cloud: a case of Dark Attractor?}

\author{Igor D.\ Karachentsev}
\address{Special Astrophysical Observatory of the  Russian Academy
  of Sciences, Nizhniy Arkhyz, Karachei-Cherkessia, 369167, Russia\\
Universite de Lyon, Univ. Lyon 1, CNRS/IN2P3, IPNL, Villeurbanne, France}
\mailto{ikar@luna.sao.ru}
\author{Olga G.\ Nasonova}
\address{Special Astrophysical Observatory of the  Russian Academy
of Sciences, Nizhniy Arkhyz, Karachei-Cherkessia, 369167, Russia\\
Universite de Nice - Sophia Antipolis, Observatoire de la C\^{o}te d'Azur
Laboratoire Cassiopee, UMR 6202, BP-4229, 06304, Nice Cedex 4, France}
\mailto{phiruzi@gmail.com}
\author{Helene M.\ Courtois}
\address{Universite de Lyon, Univ. Lyon 1, CNRS/IN2P3/INSU, IPNL, Villeurbanne, France\\
IFA, Univ. Hawaii, 2680 Woodlawn Drive, HI 96822 Honolulu, USA}
\mailto{h.courtois@ipnl.in2p3.fr}

\begin{abstract}
 We notice that nearby galaxies having high negative peculiar velocities
are distributed over the sky very inhomogeneously. A part of this
anisotropy is caused by the ``Local Velocity Anomaly'', i.e. by
the bulk motion of nearby galaxies away from the Local Void. But a
half of the fast-flying objects reside within a small region $[RA=
11.5^h - 13.0^h, Dec.= +20^{\circ} - +40^{\circ}]$, known as the
Coma~I cloud. According to Makarov \& Karachentsev (2011), this
complex contains 8 groups, 5 triplets, 10 pairs and 83 single
galaxies with the total mass of $4.7\cdot10^{13}M_{\odot}$.

We use 122 galaxies in the Coma~I region with known distances and
radial velocities $V_{LG} < 3000$ km/s to draw the Hubble relation
for them. The Hubble diagram shows a Z-shape effect of infall  with an
amplitude of +200 km/s on the nearby side and $-700$ km/s on the
back side. This phenomena can be understood as the galaxy infall
towards a dark attractor with the mass of $\sim2\cdot10^{14}M_{\odot}$ 
situated at a distance of 15 Mpc from us. The existence of large
void between the Coma and Virgo clusters affects probably the
Hubble flow around the Coma~I also.
\end{abstract}

\noindent{\it keywords/\ {galaxies: distances and redshifts --- galaxies: clusters:
individual (Coma~I)}}\\

\noindent{Accepted for publication in ApJ September, 3rd,  2011}

\section{Introduction}

Peculiar motions of galaxies relative to the uniform Hubble
expansion  can be divided into two types: 1) the virial (thermal)
motion in groups and clusters, and 2)  coherent flows (or ``Gulf
Streams'') triggered by the continuing formation of the
large-scale structure of the universe. In the first case   the
galaxies `` forget'' their initial conditions. Characteristic
amplitudes of the virial motions amount to $\sim(70-700)$ km/s at
the typical sizes of virial regions amounting to
$\sim(0.3-3)$~Mpc. In the second case, the flow amplitudes  can be also
large, reaching $\sim700$~km/s (an example --- the motion of the
Local Volume relative to the cosmic microwave background, CMB,
with the velocity of 630 km/s (Kogut et al. 1993)). It is assumed
that linear scales of the flows range from $\sim10$ to
$\sim200$ Mpc.

In a simplified view, the cosmic currents appear as a radial infall
of galaxies towards local attractors, or as radial motions
outward from the centers of cosmic voids.

If the virial motions in groups and especially in clusters are
easily measurable even at large distances, the cosmic flows are still
mostly  unstudied. The main reason is the scarcity of available
database on galaxy distances.

According to Tonry et al. (2000), our Local Group and its wider
vicinities are moving to the nearest Virgo cluster with a velocity
of $\sim180$ km/s, and to a more distant and more massive Great
Attractor in the Hydra-Centaurus with a velocity of $\sim400$
km/s. Some authors (Scaramella et al. 1995) explain the observed
motion of the Local Volume relative to the CMB by the
gravitational influence of the supercluster of galaxies, called
the Shapley Concentration, located at a distance of 200 Mpc from
us.

Based on much more numerous data on individual distances to
nearby galaxies, Tully at al. (2008) have divided the global
motion of the Local Volume into three approximately  orthogonal
components: the infall towards the Virgo Cluster with a
velocity of 185 km/s, a recession  from the expanding
Local Void with 260 km/s, and a motion towards the Centaurus
cluster with a velocity of 455 km/s.

According to the results of N-body simulations of the evolution of
large-scale structure  (Klypin et al. 2003, Schaap 2007), the
pattern of collective galaxy motions on the scales of 10--100 Mpc
can look much more intricate. Beyond the boundaries of
virialized groups and clusters, in the regions of low density
(occupying more than 90\% of the universe's volume), the
characteristic kinematic pattern appears as the motion of
galaxies from voids towards walls and filaments,
embordering them, and as flows along the filaments to
large concentrations of mass, i.e. towards the clusters. According
to the results of modeling, the amplitude of such local bulk
motions can reach up to several hundred km/s (see Figs. 5--7 in
Klypin et al. 2003).

The data on radial velocities and distances of galaxies
 in the vicinity of the closest groups and clusters: the
Local Group, M81 group, Centaurus A group, Virgo cluster,
Fornax/Eridanus cluster reveal the presence of the expected infall
zones around them (Karachentsev et al. 2007a, 2009, Karachentsev
\& Nasonova 2010, Nasonova et al. 2011).

Unfortunately, the study of collective galaxy motions in the low
density regions is not yet supported by a sufficient amount
of observational data. Only the nearest volume around us contains
a reasonable number of galaxies with measured distances in order
to get the initial idea on the ``Gulf Streams'', associated with
the walls and filaments of the large-scale structure. In our paper
we attempt to assess the observational capabilities in this
aspect.

\section{Local ($V_{LG}<600$ km/s) galaxies with high negative peculiar velocities}

Back in the 1980s, it was noted that nearby galaxies with large
negative peculiar radial velocities are distributed in the sky
extremely inhomogeneously. Except for the central region of the
Virgo cluster, negative velocities, $V_{pec}=V_{LG}-H_0\cdot
D<-300$ km/s, occur mainly in the region of the
Gemini-Cancer-Leo constellations. This phenomenon has been called
the ``Local Velocity Anomaly'' $\equiv$ LVA (de Vaucouleurs \&
Bollinger 1979, de Vaucouleurs \& Peters 1985, Giraud 1990, Hun \&
Mould 1990). The most plausible explanation for this phenomenon
was proposed by Tully et al. (1992, 2008), under which the Local
Cloud (Sheet) is receding away from the Local Void, thus
approaching the neighboring clouds in the Leo and Cancer with a
relative velocity of $\sim300$ km/s. As a result, such galaxies as
UGC3755, DDO47, KK65, D634-03, located at a distance of $7 - 9$
Mpc, possess radial velocities relative to the LG of less than 200
km/s.  Note that distances to these galaxies were
measured with high accuracy based on the Tip of Red Giant Branch (TRGB)
method with the Hubble Space Telescope (HST).

Continuing to measure distances to nearby galaxies by the TRGB
method at the HST, Karachentsev et al. (2006) drew attention to
other cases of galaxies with large negative peculiar velocities:
IC779, KK127, NGC4150, UGC7321. These galaxies are concentrated in
a small region of the sky around
$12^h10^m+30^{\circ}$, where a scattered group Coma~I is located.

The most representative sample of nearby galaxies to date, the
Catalog of Neighboring Galaxies (Karachentsev et al. 2004),
contains 450 galaxies with distance estimates within 10 Mpc from
us. If the individual distance to a galaxy was unknown, it was
included in the Local Volume sample based on the value of its
corrected radial velocity $V_{LG}<550$ km/s. The second, expanded
version of this catalog (Karachentsev, 2012, in preparation)
contains about 770 galaxies with $D\leq10$ Mpc or $V_{LG}<600$
km/s. Thirty-six galaxies with the velocities of $V_{LG}<600$ km/s, but
having the distance estimates exceeding 12 Mpc made it into this
sample too. At the Hubble parameter $H_0=73$ km/s/Mpc, these
galaxies have peculiar radial velocities of $V_{pec} <-276$ km/s.
Their list is presented in Table 1. For the obvious reason   it
does not include the galaxies in the virial zone with the radius
of 6$^{\circ}$ around the center of the Virgo cluster.

The columns of Table 1 contain: (1) galaxy name, (2) its
equatorial coordinates, (3) morphological type on  de
Vaucouleurs scale; (4) radial velocity relative to the centroid of
the Local Group; (5) distance estimate in Mpc; (6) the method used
for distance measurement (``tf''--- based on the Tully-Fisher
relation with the parameters from Tully et al. (2008), ``sbf''
--- based on the surface brightness fluctuations from Tonry et al. (2001),
``bs'' --- based on the luminosity of the brightest stars); (7)
apparent B-band magnitude; (8) doubled amplitude of
rotation in km/s; (9) peculiar radial velocity $V_{LG}-73\cdot D$
in km/s; (10) comments on probable galaxy membership in
certain groups. Most data on magnitudes and rotational velocities are taken
from NED and HyperLEDA databases and supplemented with Courtois et al. (2009,
2011) and Cannon et al. (2011).

Figure~1 shows the distribution of 8811 galaxies in the sky in
equatorial coordinates  with radial velocities $V_{LG}<3000$ km/s
at the galactic latitudes $\mid b\mid> 15^{\circ}$.  The gray
ragged strip indicates the zone of strong galactic
absorption according to Schlegel et al. (1998). The figure clearly
shows the filamentary structure of the Local Supercluster and its
immediate environs. The red circles mark 36 galaxies from Table 1
with large negative $V_{pec}$. The distribution of the  36 marked
galaxies is characterized by a significant clumpiness, as well as
by the west-east asymmetry. A part of this asymmetry is caused by
the LVA phenomenon, caused by the runaway of galaxies  from a
large volume of the Local Void centered approximately at $[19.0^h,
+3^{\circ}]$. The median peculiar velocity of the ``flyer''
galaxies  is 660 km/s. With a median distance of the galaxies
amounting to 15.1 Mpc, a typical error of distance measurement
using the Tully-Fisher method (about 20\%) generates a
characteristic error of peculiar velocity amounting to $\sim220$
km/s. Consequently, some galaxies could appear in Table 1 due to
the distance measurement errors. Another reason could be the entry
of galaxies in a group or a cluster with large virial velocities.
Some cases like this are commented in the last column of Table 1. In particular, the case
of an E-galaxy NGC1400, belonging to the Eridanus group, was
already discussed by Trentham et al. (2006). Curiously, this list
includes the galaxy UGC12713 = LOG506, which is included in the
catalog of 513 most isolated objects of the Local Supercluster
(Karachentsev et al. 2011).

Exactly one half of the galaxies in Table 1 is concentrated within the
region $RA=[11.5^h, \- 13.0^h]$,
$Dec.[+20^{\circ},+40^{\circ}]$, occupying only 1\% of the total
area of the sky. The probability of random formation of such a
configuration is vanishingly small   $(\sim10^{-30})$. This
subsystem of ``flyers'', designated in Table 1 as Coma~I cloud, is
characterized by the median distance of 16.2 Mpc and the median
peculiar velocity of $-740$ km/s. A complex kinematic situation in
this region, which lies on the equator of the Local Supercluster,
deserves a more detailed examination.

\section{The Coma~I complex of galaxies}

According to Tully (1988) in the region  $RA=[11.5^h,
13.0^h], Dec.=[+20^{\circ}, +40^{\circ}$], outlined by a red
contour in Fig.\,1, there is a group of galaxies ``14--1'',
belonging to the  Coma-Sculptor cloud. From 25 galaxies with known
radial velocities Tully has determined the mean radial velocity of
the group as +911 km/s and its total luminosity of
$L_B=1\cdot10^{11}L_{\odot}$. The virial radius of the group
according to Tully is 0.34 Mpc, the dispersion of radial
velocities   $\sigma_v=266$  km/s and the ratio of virial mass to
luminosity $M_{VIR}/L_B=523M_{\odot}/L{\odot}$.

At present, 206 galaxies are known in this region  with  radial
velocities $V_{LG}<3000$  km/s. The distribution of these
galaxies on the velocity scale is shown in Fig.\,2. The early-type
galaxies with developed bulges ($T<3)$ are marked in gray. A small
number of galaxies with radial velocities $V_{LG}>1200$ km/s
catches one's eye. This is probably due to the presence of a large
cosmic void between the Local Supercluster and the rich Coma cluster
(+6900 km/s).

Some authors made a search for dwarf galaxies in the region of
Coma~I. Almost a half of this region has been inspected by
Binggeli et al. (1990)  on the specially exposed IIIaJ Palomar
Schmidt plates, resulting in to the discovery of 34 dwarf
galaxies. Trentham \& Tully (2002) used the 8-m Subaru telescope
to search for  the dwarf population in Coma~I in a narrow strip near
NGC4274. In the area of 1.3 square degrees, they have detected
about 30 dwarf member candidates of the Coma~I group. Karachentsev
et al. (2007b) found in the region  $RA=[11.5^h,
13.0^h], Dec.=[+20^{\circ},+40^{\circ}$] twenty-five dwarf
galaxies of low surface brightness. Subsequent observations in the
HI 21cm line (Huchtmeier et al. 2009) added several new dwarf
members of the Coma~I complex. About 12\% of the studied region
overlap with the zone of the blind HI survey (Kovac et al. 2009),
where several new nearby gas-rich dwarfs were discovered. The SDSS
survey contains a lot of valuable data about the galaxies in
Coma~I (Abazajian et al. 2009). Boselli \& Gavazzi (2009) used the
HI characteristics of the galaxies in this region to search for
the effect of "HI-deficiency".

Makarov \& Karachentsev (2011) collected the observational data on
10 800 galaxies in the Local Supercluster and its vicinity with
$V_{LG}< 3500$ km/s and $\mid b\mid>15^{\circ}$. A new clustering
algorithm was applied to this sample, which takes into account
individual luminosities, radial velocities and mutual distances of
galaxies. In contrast to the simple friend-of-friend percolation
algorithm (Huchra \& Geller 1982, Crook er al. 2007), a new
criterion isolates the galaxy systems with approximately the same
characteristics both in close and distant volumes. In the examined
region  $RA=[11.5^h, 13.0^h],
Dec.=[+20^{\circ},+40^{\circ}$] eight groups (Makarov \&
Karachentsev 2011), 5 triplets (Makarov \& Karachentsev 2009) and
10 pairs of galaxies (Karachentsev \& Makarov 2008) were isolated
with a total  of $n_v =122$ members. Apart from them, 83 galaxies
have evaded the  clustering, representing 41\% of the total. The
data on the group members, the members of triplets and pairs, and
single galaxies are listed in Tables 2, 3 and 4, respectively. The
columns of tables contain: (1) galaxy name, (2) its coordinates,
(3) radial velocity relative to the $LG$, (4) morphological type, (5)
the apparent $K_s$-magnitude, (6,7) the distance to the galaxy,
indicating the method of distance measurement: ``cep'' --- from
the luminosity of Cepheids, ``rgb''
--- from the luminosity of the tip of red giant branch, ``sbf''--- from the
surface brightness fluctuations , ``tf'' --- by the Tully-Fisher
method.  The assumed typical errors of these methods are:
7\% for "cep" and "rgb", 12\% for "sbf", and 20\% for "tf".

The distribution of 206 galaxies of different degrees of
multiplicity is presented in equatorial coordinates in Fig.~3.
Early-type galaxies ($T <$3) are marked in gray. The members of
the eight groups are indicated by squares, the members of triplets
and pairs --- by triangles, and single objects are shown by small
circles. The larger symbols mark bright galaxies with $m_K<8.0^m$.
Note that among 83 field galaxies there are only two early-type
galaxies: CGCG186-009 and UGC7816, and objects of high luminosity
are almost absent.

As we can see, there exists a mutual overlap of members of
different systems in this sky region. However, this situation is
absolutely expectable, since the region is located on the very
equator of the Local Supercluster.

The summary of the main characteristics of the eight MK-groups,
forming the Coma~I complex is presented in Table 5, its columns
contain: (1) the name of the brightest galaxy of the group, (2)
the number of members with measured radial velocities, (3) the
mean radial velocity based on which the distance to the group was
 determined at $H_0=73$ km/s/Mpc, (4) radial velocity dispersion, (5)
linear harmonic radius of the group in kpc, (6) the total
luminosity of the group members in the $K_s$-band, (7.8) virial
mass and virial mass ratio to the total K-luminosity, (9) the
number of group members with measured distances, (10,11) the mean
distance modulus of the group and its variance. In the lower rows
of columns (5--8) we list the values of group parameters, not
derived by the average velocity of the group, but from the
average of individual distance moduli of the group members.

It follows from these data that the groups of galaxies have a
typical linear size of $\sim300$ kpc and a rather low dispersion
of radial velocities. Only one group, NGC4150, stands out by a
very high virial mass-to-luminosity ratio,
$M_{VIR}/L_K=830M_{\odot}/L_{\odot}$.  However, this value drops
by a factor of 5 when the distance to the group is estimated
directly, rather than by the mean radial velocity.

Most of the individual distance moduli for galaxies in this
complex were measured from the Tully-Fisher relation with a
typical error of $\sim0.4$ mag. The values of $\sigma(DM)$ in the
last column of Table 5 show that 7 out 8 groups may be regarded as
real physical groups.
Only in the case of group NGC4631 the applied clustering algorithm
isolates a pseudo-group of galaxies with close
radial velocities, but different distance estimates (subgroups
around NGC4631, NGC4278 and NGC4414).  Despite the small
statistics, a good agreement between the distance estimates is as
well observed in the pairs and triplets of galaxies (see Table
3).

\section{Hubble flow in the Coma~I complex}

A group of galaxies around NGC~4150 looks like a ``flock'' heading
towards us with an average velocity of $-1002$ km/s relative to
the uniform Hubble expansion. Its angular distance from the center
of the Virgo cluster is 17$^{\circ}$, what is smaller than the
zero-velocity surface radius of the cluster,  23$^{\circ}$
(Karachensev \& Nasonova 2010). At the first sight, a high
peculiar velocity of NGC4150 can be explained by its participation
in the Virgocentric infall. However, the group is located at a
distance of 16.3 Mpc from us, almost as far as the center of the
Virgo  cluster (16.8 Mpc). Therefore, the radial infall of the
NGC4150 group toward the center of the cluster should not produce a
significant component of the line-of-sight velocity. Consequently,
there exists another reason of high peculiar velocity in this
group as a whole.

The distribution of 122 galaxies in the Coma~I region by radial
velocities and distance estimates is presented in Fig.~4. 
Its upper and lower pannels show the same observational data but
given in two different manner: in the linear scale and in a log-log
view. Different symbols denote galaxies of different morphology and
membership in the same manner as in Fig.3. Distance error bars are
marked by light horizontal lines.
The straight line there corresponds to the uniform Hubble flow with
the parameter of $H_0=73$ km/s/Mpc. The broken line shows the
behavior of the running median with a window of 2 Mpc.
Notwithstanding a significant scatter of observational data, the
mean radial velocity variation with distance demonstrates the
well-known effect of a Z-shaped wave that results from the infall
of galaxies towards a local attractor. The characteristic infall
velocity at the front side is about 200 km/s, and
at the back side it amounts to about 700 km/s. Two dashed lines in
Fig.~4 show the order of infall on the attractor with the mass of
$0.5\cdot10^{14}$ and $2.0\cdot10^{14}M_{\odot}$, located at a
distance of 15~Mpc from us, for the line of sight, crossing the
center of the attractor. A comparison of these curves with the
running median suggests that the mass of the hypothetical attractor
is $\sim2\cdot10^{14}M_{\odot}$.

However, the distribution of galaxies in Fig.~1 does not reveal
any prominent concentration of galaxies in the region of  Coma~I.
In other words, what we have here is an unusual case of a Dark
Attractor. The total virial mass of 8 groups from Table~5 is
$3.5\cdot10^{13}M_{\odot}$. Adding to it the total mass of five
triplets ($0.3\cdot10^{13}M_{\odot}$), 10 pairs
($0.2\cdot10^{13}M_{\odot}$) and single galaxies
($\sim0.7\cdot10^{13}M_{\odot}$), we derive the total mass of the
Coma~I  complex of about ($4.7\cdot10^{13}M_{\odot}$). This value
is four times smaller than the expected mass to explain the infall
amplitude in Fig.~4.

As noted by Makarov \& Karachentsev (2011), the total mass of the
dark matter concentrated in the virial regions of groups and
clusters corresponds to the average density
$\Omega_m=0.08\pm0.02$, which is three times lower than the global
density of $0.28\pm0.03$ (Spergel et al. 2007). An assumption that
2/3 of the amount of dark matter in the universe is located
outside the virialized regions can be one of the explanations of
this contradiction. Massive, but unvirialized filaments and walls
of the large-scale structure may appear to be considerable
reservoirs of dark matter as well as ``lost'' baryons (Fukugita \&
Peebles 2004). It is quite likely that the region of Coma~I is a
fragment of such a massive dark filament, extending far to the
north of the Virgo cluster.

\section{Concluding remarks}

The detection of galaxies or galaxy groups, moving with peculiar
velocities of $\sim(500-1000)$ km/s is a sophisticated
observational task. Such motions in the low density regions,
comparable in amplitude with virial velocities in clusters,
probably arise from the continuing formation of elements of the
large-scale structure, namely, filaments and walls.

Large peculiar motions in the  Coma~I cloud are not related with a
high concentration of galaxies in this field. The observed
relationship between the velocities and distances in the Coma~I
resembles the pattern of galaxy infall towards an invisible Dark
Attractor. Such an attractor can be a dark filament ($\sim200$
galaxies + dark matter), which extends to the north of the Virgo
cluster. Behind this filament there is a large cosmic void, lying
between the Virgo and Coma~I clusters. It is possible that the
presence of this void causes the three times
greater infall amplitude at the far side of the filament in the
Coma~I compared to the front side. The observational resources
for the investigations of motions in Coma~I are far from being
exhausted. The scatter of galaxies on the Hubble diagram (Fig. 4)
is partly due to the low quality of the available HI data, used to
determine the distances by the Tully-Fisher method. For example,
according to the LEDA, the galaxy IC2992=Mrk757 has the HI
line width of $W_{50}=138$ km/s, whereas according to the recent
observations at the Arecibo, it amounts to only 46 km/s (R.
Giovanelli, personal communication). Within the SDSS survey
(Abazajian et al. 2009)  a large number of dwarf galaxies were
detected in Coma~I, in which the HI line widths have not yet been
measured. Almost entire region of the Coma~I is situated in the
ALFALFA survey zone (Giovanelli et al. 2005). New data from this
survey can render the pattern of kinematics in the Coma~I galaxies
much more clear \footnote{The new ALFALFA data release (Haynes et al, arXiv:1109.0027) yields
at least two new distant dIr galaxies with low velocities in the 
Coma I field: AGC229053 ( D= 17.9 Mpc, V$_{LG}$ = +376 km/s) and
AGC749236 ( D= 19.5 Mpc, V$_{LG}$ = +235 km/s).}.

One pessimistic deduction follows from the above. For the vast
majority  of galaxies the only distance estimates available as yet
are their radial velocities. To account for non-Hubble motions of
galaxies, and thereby improve the accuracy of kinematic distance
measurement $V/H_o$, various models correcting for the local
galaxy infall towards the Virgo attractor and the Great Attractor
in the Centaurus have been proposed (Kraan-Korteweg 1986, Masters
2005). Such models may, however, turn out to be too naive, not
reflecting the true pattern of large-scale flows.

\ack
{I.D. Karachentsev thanks Riccardo Giovanelli for the provision of
the ALFALFA data prior to publication. This work was partially
supported  by the following grants: the Russian Foundation for
Basic Research (grant no.~11--02--00639), the CNRS, a grant from
the Space Telescope Science Institute under the NASA contract
within GO12546, a grant of the Ministry of Education and Science
of the Russian Federation N 14.740.11.0901. O.G. Nasonova thanks
the non-profit Dmitry Zimin's Dynasty Foundation for the financial
support.}

\section*{References}
\begin{harvard}

\item[Abazajian K.N., Adelman-McCarthy J.K., Agueros M.A. et al., 2009, ApJSuppl. 182, 543]
\item[ Binggeli B., Tarenghi M., Sandage A., 1990, A \& A, 228, 42]
\item[ Boselli A., Gavazzi G., 2009, A \& A, 508, 210]
\item[ Cannon J.M., Giovanelli R., Haynes M.P. et al., 2011, ApJL, in press
(ArXiv:1105.4505) ]
\item [Courtois H.M., Tully R.B., Makarov D.I. et al., 2011, MNRAS, 411, 2005]
\item[  Courtois H.M., Tully R.B., Fisher J.R. et al., 2009, AJ, 138, 1938]
\item[ Crook A.C., Huchra J.P., Martimbeau N. et al., 2007, ApJ, 655, 790]
\item[ de Vaucouleurs G., Peters W.L., 1985, ApJ, 297, 27]
\item[ de Vaucouleurs G., Bollinger G., 1979, ApJ, 233, 433]
\item[ Fukugita M., Peebles P.J.E., 2004, ApJ, 616, 643]
\item[ Giovanelli R., Haynes M.P., Kent B.R., et al. 2005, AJ, 130, 2598]
\item[ Giraud E., 1990, A\&A, 231, 1]
\item[ Han M., Mould J., 1990, ApJ, 360, 448]
\item[ Huchra J.P., Geller M.J., 1982, ApJ, 257, 423]
\item[ Huchtmeier W.K., Karachentsev I.D., Karachentseva V.E., 2009, A \& A, 506, 677]
\item[ Karachentsev I.D., Makarov D.I., Karachentseva V.E., Melnyk O.V., 2011,
      Astrophys. Bulletin, 66, 1]
\item[ Karachentsev I.D., Nasonova O.G., 2010, MNRAS, 405, 1075]
\item[ Karachentsev I.D., Kashibadze O.G., Makarov D.I., Tully R.B., 2009, MNRAS,
      393, 1265]
\item[ Karachentsev I.D., Makarov D.I., 2008, Astrophys. Bulletin, 63, 299]
\item[ Karachentsev I.D., Karachentseva V.E., Huchtmeier W.K., 2007a, Astron.Lett., 33, 512]
\item[ Karachentsev I.D., Tully R.B., Dolphin A. et al. 2007b, AJ, 133, 504]
\item[ Karachentsev I.D., Dolphin A.E., Tully R.B., et al. 2006, AJ, 131, 1361]
\item[ Karachentsev I.D., Karachentseva V.E., Huchtmeier W.K., Makarov D.I.,
               2004, AJ, 127, 2031 (CNG)]
\item[ Klypin A., Hoffman Y., Kravtsov A.V., Gottloeber S., 2003, ApJ, 596, 19]
\item[ Kogut A., Lineweaver C., Smoot G.F., et al., 1993, ApJ, 419, 1]
\item[ Kovac K., Oosterloo T.A., van der Hulst J.M., 2009, MNRAS, 400,743]
\item[ Kraan-Korteweg R.C., 1986, A \& AS, 66, 255]
\item[ Makarov D.I., Karachentsev I.D., 2011, MNRAS, 412, 2498]
\item[ Makarov D.I., Karachentsev I.D., 2009, Astrophys. Bulletin, 64, 24]
\item[ Masters K.L., 2005, ``Galaxy flows in and around the Local Supercluster''q, PhD, Cornell Univ.]
\item[ Nasonova O.G., de Freitas Pacheco J.A., Karachentsev I.D., 2011, A\&A, 532, A104]
\item[ Scaramella R., 1995, ApL \& C, 32, 297]
\item[ Schaap W., 2007, PhD Thesis, Groningen Univ.]
\item[ Schlegel D.J., Finkbeiner D.P., Davis, M., 1998, ApJ, 500, 525]
\item[ Spergel D.N., et al. 2007, ApJS, 170, 377]
\item[ Tonry J.L., Dressler A., Blakeslee J.P., et al. 2001, ApJ, 546, 681]
\item[ Tonry J.L., Blakeslee J.P., Ajhar E.A., Dressler A., 2000, ApJ, 530, 625]
\item[ Trentham N., Tully R.B., Mahdavi A., 2006, MNRAS, 369, 1375]
\item[ Trentham N., Tully R.B., 2002, MNRAS, 335, 712]
\item[ Tully R.B., Shaya E.J., Karachentsev I.D., et al., 2008, ApJ, 676, 184]
\item[ Tully R.B., Pierce M.J., 2000, ApJ, 533, 744]
\item[ Tully R.B., Shaya E.J., Pierce M.J., 1992, ApJS, 80, 479]
\item[ Tully R.B., 1988, Nearby Galaxies Catalog, Cambridge University Press]
\end{harvard}

\newpage
\begin{figure}
\includegraphics[width=0.6\textwidth,angle=-90]{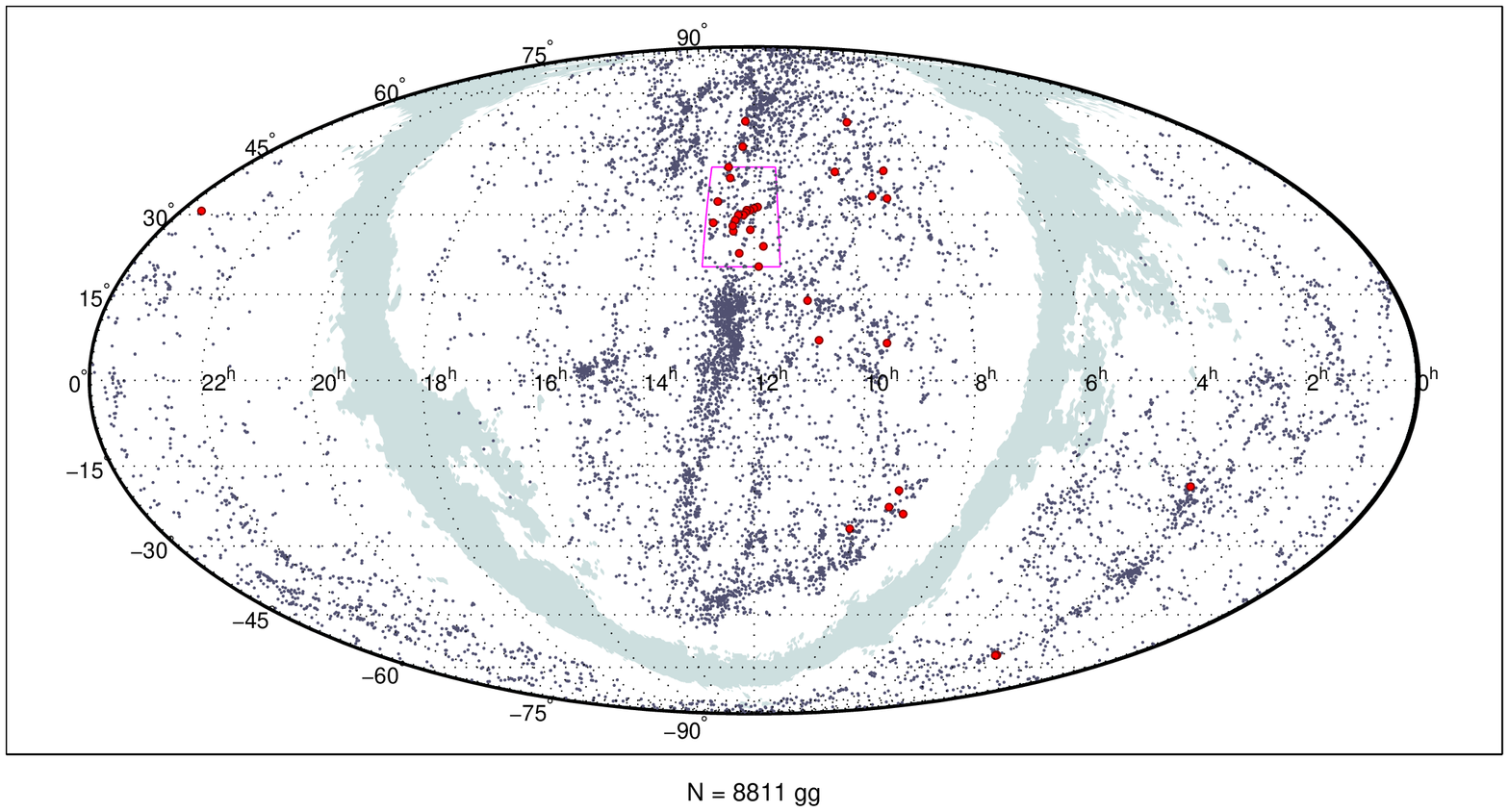}
\caption{The distribution of 8811 galaxies (marked with points)
with radial velocities of $V_{LG} <3000$ km/s in the sky in
equatorial coordinates. The circles mark the galaxies with large
negative peculiar velocities. A half of them is concentrated in
the Coma~I region inside the rectangular contour. The region of
strong absorption along the Milky Way is filled in gray.}
\end{figure}

\begin{figure}
\includegraphics[width=0.7\textwidth,angle=-90]{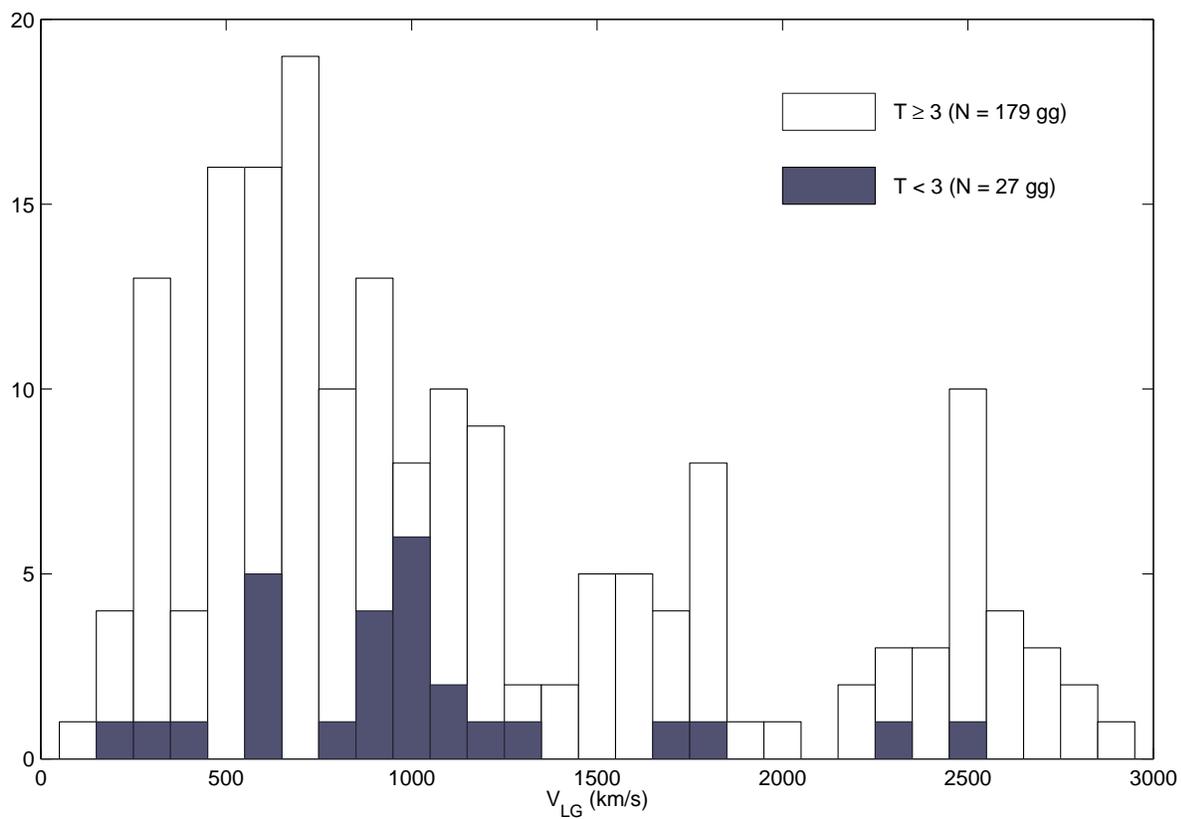}
\caption{The distribution of 206 galaxies in the Coma~I cloud by
radial velocities. The early-type galaxies of $T <$3 are shaded.}
\end{figure}

\begin{figure}
\begin{center}
\includegraphics[width=0.7\textwidth,angle=-90]{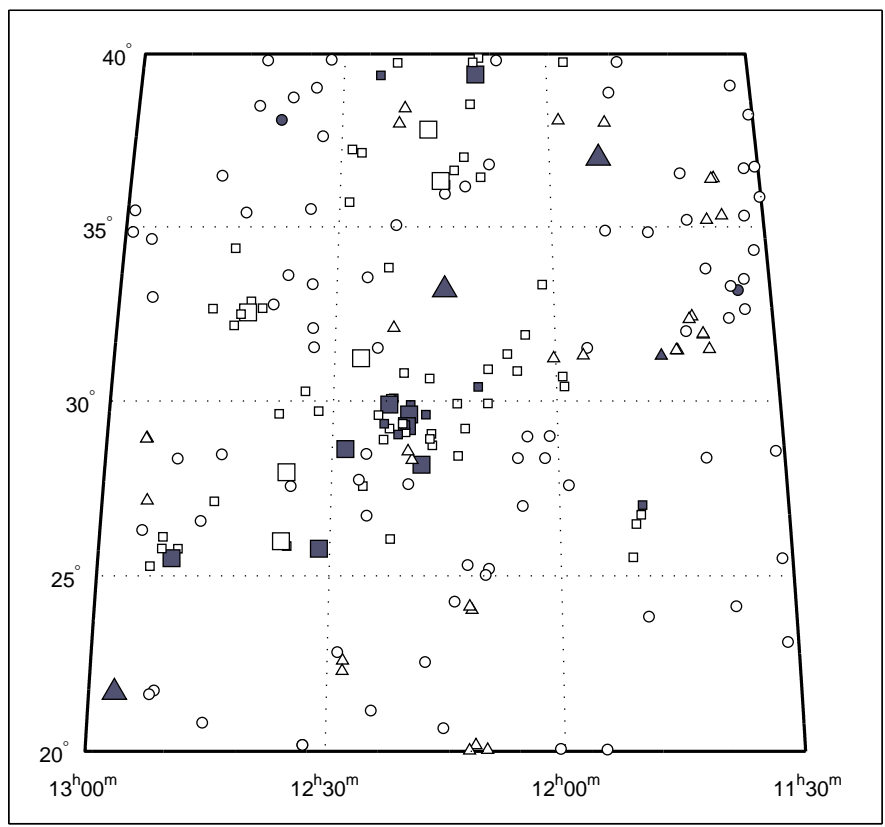}
\end{center}
\caption{The distribution of 206 galaxies in the Coma~I region on
a larger scale. The squares mark the group members, the triangles
--- the members of triplets and pairs, the circles --- single galaxies.
The galaxies with bulges ($T<$3) are shown in gray. Eighteen
brightest galaxies with $m_K<8.0^m$ are marked by larger symbols.}
\end{figure}

\begin{figure}
\begin{center}
\begin{tabular}{c}  
\includegraphics[width=0.6\textwidth,angle=-90]{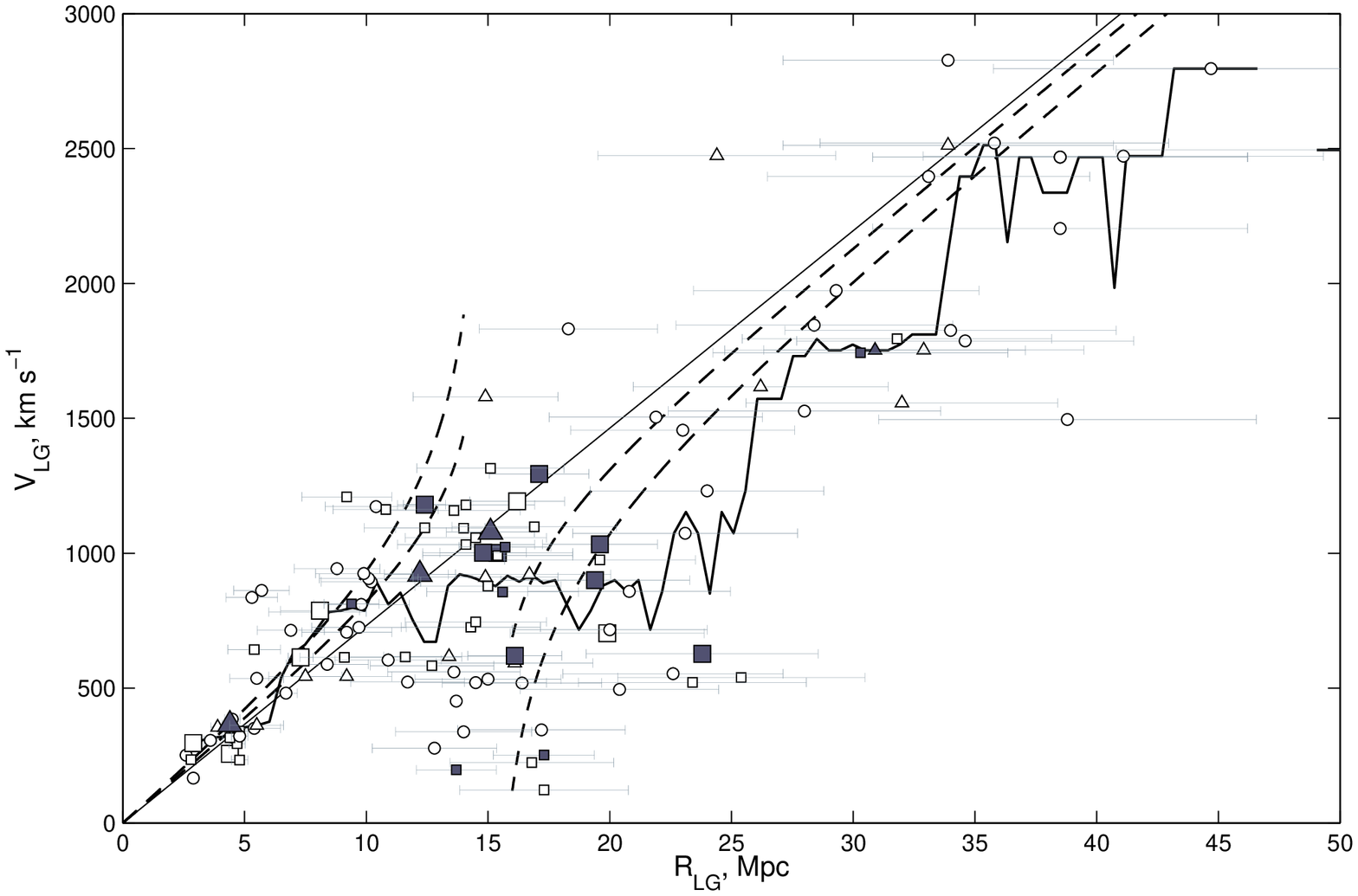}\\
\includegraphics[width=0.6\textwidth,angle=-90]{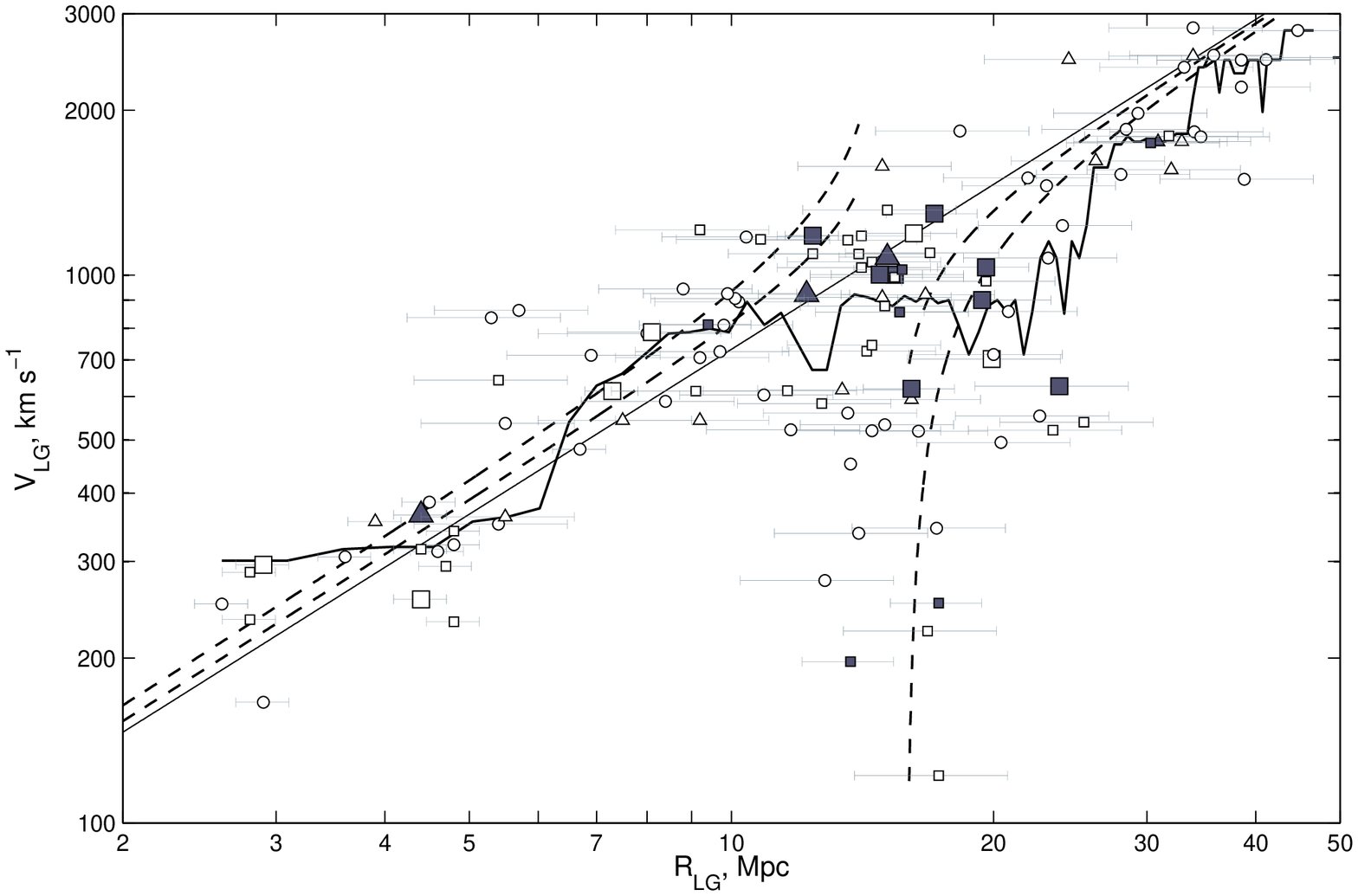}
\end{tabular}
\end{center}
\caption{The velocity--distance relation for 122 galaxies in the
Coma~I region shown in the linear (top) and a log-log (bottom)
presentations.
 The straight line corresponds to the unperturbed Hubble flow with
the parameter $H_0=73$ km/s/Mpc. The symbols, marking different galaxies
are the same as in Fig.3. Distance error bars are shown. The broken
line demonstrates the behavior of the running median with a window of
2 Mpc. Two dashed lines correspond to the pattern of the galaxy
infall towards a point-like attractor with a mass of
  $0.5\cdot10^{14}M_{\odot}$  and
$2.0\cdot10^{14}M_{\odot}$, located at a distance of 15 Mpc (where
the line of sight passes through the attractor's center).}
\end{figure}

\renewcommand{\baselinestretch}{0.8}
\begin{table}
\caption{List of 36 galaxies with $V_{LG} < 600$ km/s and $D > 12$ Mpc
		(outside the Virgo cluster core)}
\begin{tabular}{lcrccllrrl} \\ \hline

Name      &     RA    Dec   &  T&$V_{LG}$&  $D$   & $[D]$ &  $B_t$ & 2$V_m$&  $V_{pec}$&  Notes          \\
    &     (2000.0)    &   &km/s&  Mpc &     &  mag &km/s& km/s  &                 \\
\hline
  (1)     &    (2)          & (3)& (4)   & (5)     & (6)    & (7)     & (8)   &   (9)    & (10)                \\
\hline
NGC1400   & 033930.8$-$184117 &-3 & 485& 24.5 &  sbf& 11.92&  --- &  $-$1303 &  Eridanus gr    \\
IC 2038   & 040854.1$-$555932 & 7 & 505& 17.0 &  tf & 14.98&  95&  $-$736 &  pair w. N1533  \\
NGC1533   & 040951.8$-$560706 & 1 & 582& 19.4 &  sbf& 11.79& 239&  $-$834 &  pair w. IC2038 \\
UGC4704   & 085900.3$+$391236 & 8 & 581& 15.2 &  tf & 15.0 &  96&  $-$529 &  LVA            \\
ESO497-04 & 090303.1$-$234830 & 8 & 519& 14.5 &  tf & 16.36&  78&  $-$540 &                 \\
UGC4787   & 090734.9$+$331636 & 8 & 490& 15.1 &  tf & 14.6 & 106&  $-$612 &  LVA            \\
UGCA153   & 091312.1$-$192431 & 8 & 488& 19.0 &  tf & 15.40&  96&  $-$899 &                 \\
UGC4932   & 091934.1$+$510633 & 8 & 598& 18.6 &  tf & 15.17& 100&  $-$760 &  N2841 gr?      \\
DDO62     & 092127.5$-$223002 & 8 & 592& 14.4 &  tf & 14.8 & 103&  $-$459 &                 \\
SDSS      & 092609.4$+$334304 &10 & 488& 19.7 &  tf & 17.8 &  47&  $-$950 &  LVA            \\
CGCG035-07& 093444.9$+$062532 & 9 & 356& 13.0 &  tf & 15.22&  92&  $-$593 &  LVA?           \\
ESO499-38 & 100350.2$-$263646 &10 & 591& 12.7 &  tf & 15.68&  68&  $-$336 &                 \\
KKH58     & 100722.7$+$385811 & 8 & 569& 14.9 &  tf & 15.61&  72&  $-$519 &  LVA?           \\
UGC5923   & 104907.6$+$065502 & 7 & 537& 16.4 &  tf & 14.4 & 122&  $-$660 &  Leo I gr       \\
NGC3489   & 110018.6$+$135404 & 1 & 538& 12.1 &  sbf& 11.12& 282&  $-$345 &  Leo I gr       \\
DDO97     & 114857.2$+$235016 &10 & 453& 13.7 &  bs & 15.14&  70&  $-$547 &  Coma I cld     \\
UGC6881   & 115444.7$+$200320 &10 & 519& 16.4 &  tf & 15.9 &  72&  $-$678 &  Coma I cld     \\
KDG82     & 115539.4$+$313110 & 8 & 560& 13.6 &  tf & 14.84&  87&  $-$433 &  Coma I cld     \\
KUG1157+31& 120016.2$+$311330 & 8 & 593& 16.1 &  tf & 15.05&  91&  $-$582 &  Coma I cld     \\
NGC4080   & 120451.8$+$265933 & 8 & 519& 15.0 &  tf & 13.7 & 134&  $-$576 &  Coma I cld     \\
IC 2992   & 120515.9$+$305120 & 9 & 583& 12.7 &  tf & 15.02&  46&  $-$344 &  Coma I cld     \\
UGC7131   & 120911.8$+$305424 & 8 & 226& 16.8 &  tf & 15.1 & 102&  $-$1000 &  Coma I cld     \\
NGC4150   & 121033.6$+$302406 &$-$1 & 198& 13.7 &  sbf& 12.49&   ---&  $-$802 &  Coma I cld     \\
KK127     & 121322.7$+$295518 & 9 & 122& 17.3 &  tf & 15.6 &  82&  $-$1141 &  Coma I cld     \\
UGC7267   & 121523.6$+$512060 & 8 & 550& 12.9 &  tf & 14.79&  95&  $-$392 &  N4157 gr?      \\
UGC7320   & 121728.5$+$444841 & 8 & 582& 14.3 &  tf & 15.4 &  72&  $-$462 &  N4258 gr       \\
UGC7321   & 121734.0$+$223225 & 7 & 344& 17.2 &  tf & 14.09& 188&  $-$912 &  Coma I cld     \\
IC 779    & 121938.7$+$295300 &$-$1 & 196& 17.3 &  sbf& 15.22&  --- &  $-$1067 &  Coma I cld     \\
IC 3247   & 122314.0$+$285338 & 7 & 539& 25.4 &  tf & 15.25& 137&  $-$1315 &  Coma I cld     \\
IC 3308   & 122517.9$+$264253 & 8 & 277& 12.8 &  tf & 15.41& 106&  $-$657 &  Coma I cld     \\
IC 3341   & 122623.4$+$274444 & 9 & 338& 14.0 &  tf & 16.4 &  65&  $-$684 &  Coma I cld     \\
UGC7699   & 123248.0$+$373718 & 7 & 508& 14.5 &  tf & 13.3 & 162&  $-$550 &  Coma I cld     \\
UGC7774   & 123622.5$+$400019 & 7 & 553& 22.6 &  tf & 14.6 & 159&  $-$1097 &  Coma I cld     \\
FGC1497   & 124700.6$+$323905 & 8 & 496& 23.4 &  tf & 16.8 &  72&  $-$1212 &  Coma I cld     \\
UGC7990   & 125027.2$+$282110 &10 & 495& 20.4 &  tf & 16.2 &  74&  $-$994 &  Coma I cld     \\
UGC12713  & 233814.5$+$304233 & 7 & 578& 12.2 &  tf & 14.81&  85&  $-$313 &  LOG506         \\
\hline
\end{tabular}
\end{table}

\begin{table}
\caption{MK- groups with $V_{LG} < 3000$ km/s within  $RA=[11.5, 13.0h], Dec.=[+20^{\circ}, +40^{\circ}$]}
\begin{tabular}{lcrrrrl}\\ \hline
    Gal    &          RA,Dec.&        $V_{LG}$& $ T$& $K_s$&  $D$& $[D]$  \\
\hline
    (1)    &           (2)   &          (3)&    (4)&   (5) &  (6)&   (7)  \\
\hline

\multicolumn{7}{c}{Group N3990}   \\
\hline
NGC3900       &J114909.5+270119 &   1743 &  1&   8.69 & 30.3 & tf   \\
UGC06791      &J114923.7+264430 &   1795 &  7&  11.32 & 31.8 & tf   \\
NGC3912       &J115004.4+262844 &   1725 &  3&   9.94 &   ---  &      \\
KUG1148+258A &J115039.9+253134 &   1748 &  7&  14.10 &  ---  &        \\
\hline
\multicolumn{7}{c}{Group N4062}\\
\hline
NGC4020       &J115856.9+302449 &    726&   6&  10.75  &14.3 & tf  \\
IC2984        &J115907.5+304151 &    683&   9&  13.27  &  ---  &    \\
UGC07007      &J120133.2+332029 &    758&   9&  13.88  &  ---  &     \\
NGC4062       &J120403.8+315345 &    745&   5&   8.20  &14.5 & tf  \\
\hline
\multicolumn{7}{c}{Group N4150}\\
\hline
UGC07131    &  J120911.8+305424 &    224 &  8 & 13.18  &16.8&  tf  \\
NGC4150     &  J121033.7+302406 &    197 & $-$2 &  8.98  &13.7&sbf \\
KK127       &  J121323.6+295519 &    122 &  9 & 13.99  &17.3&  tf  \\
IC0779      &  J121938.8+295259 &    252 & $-$1 & 11.79  &17.3&sbf \\
\hline
\multicolumn{7}{c}{Group N4151}    \\
\hline
KUG1154+400 & J115725.3+394552 &   1019 &  9  & 12.46 &   ---  &    \\
SDSS & J120002.4+424723 &   1031 &  1  & 14.72 &   ---  &           \\
SDSS & J120751.6+413347 &   1098 &  9  & 15.58 &   ---  &           \\
SDSS & J120824.5+412405 &    955 &  9  & 14.84 &   ---  &           \\
UGC07129     & J120855.1+414427 &    950 &  2  & 10.56 &   ---  &   \\
NGC4143      & J120936.1+423203 &   1002 & $-$2  & 7.85& 14.8 &sbf\\
UGC07146     & J120949.2+431405 &   1101 &  7  & 14.00 &   ---  &   \\
NGC4145      & J121001.5+395302 &   1032 &  7  &  8.47 & 14.1 & tf\\
NGC4151      & J121032.6+392421 &   1001 &  2  &  7.37 & 15.4 & tf\\
UGC07175     & J121055.4+394524 &   1173 &  8  & 12.52 &  ---  &    \\
KKH076      & J121121.8+383228 &   1094 & 10  & 15.34 & 12.4 & tf \\
UGC07207     & J121219.2+370049 &   1057 &  8  & 12.56 & 14.5 & tf\\
PGC166134    & J122207.0+394442 &   1098 & 10  & 14.66 & 16.9 & tf\\
NGC4369      & J122436.2+392259 &   1053 &  1  &  8.91 &   ---  &   \\
\hline
\end{tabular}
\end{table}

\begin{table}
\begin{tabular}{lcrrrrl}\\ \hline
    (1)    &           (2)   &          (3)&    (4)&   (5) &  (6)&   (7) \\
\hline
 \multicolumn{7}{c}{Group N4244} \\
\hline
KUG1207+367 & J120956.5+362604 &    341 & 10 & 14.12  & 4.83& rgb   \\
NGC4190      & J121344.7+363803 &    235 &  9 & 12.46  & 2.82& rgb  \\
UGCA276     & J121457.9+361308 &    287 & 10 & 12.96  & 2.86& rgb   \\
NGC4214      & J121539.2+361937 &    296 &  9 &  7.90  & 2.94& rgb  \\
NGC4244      & J121729.7+374826 &    256 &  6 &  7.72  & 4.49& rgb  \\
UGC07559     & J122705.2+370833 &    233 & 10 & 11.75  & 4.87& rgb  \\
UGC07599     & J122828.6+371401 &    294 &  8 & 12.42  & 4.70& rgb  \\
UGC07605     & J122838.9+354303 &    316 & 10 & 12.50  & 4.43& rgb  \\
\hline
\multicolumn{7}{c}{Group N4274}\\
\hline
NGC4173      & J121221.4+291225 &   1093&   7&  10.65 & 14.0 & tf  \\
PGC166125    & J121319.1+282522 &    991&  10&  16.42 & 15.4 & tf  \\
UGC07300     & J121643.3+284350 &   1179&  10&  12.64 & 14.1 & tf  \\
CS1008      & J121651.4+290318 &   1100&   8&  15.29 &   ---  &      \\
MAPS-NGP    & J121703.4+285451 &    810&   8&  15.50 &   ---  &      \\
NGC4245      & J121736.8+293629 &    856&   1&   8.30 & 15.6 & tf  \\
NGC4251      & J121808.3+281031 &   1033&  $-$2&   7.72 & 19.6 &sbf\\
NGC4274      & J121950.6+293652 &    900&   2&   7.02 & 19.4 & tf  \\
NGC4283      & J122020.8+291839 &   1023&  $-$5&   9.03 & 15.7 &sbf\\
KUG1218+310 & J122035.9+304757 &    988&   8&  13.95 &   ---  &      \\
IC3215       & J122210.4+260307 &    975&   7&  12.54 & 19.6 & tf  \\
NGC4310      & J122226.3+291231 &    878&   3&   9.67 & 15.0 & tf  \\
NGC4314      & J122232.0+295343 &    959&   1&   7.44 &   ---  &      \\
VV279b      & J122754.0+283801 &   1024&  10&  13.13 &   ---  &        \\
\hline
\multicolumn{7}{c}{Group N4565} \\
\hline
NGC4494       &J123124.0+254630 &   1294 & -5 &  6.79 & 17.1 & sbf  \\
NGC4525       &J123351.2+301639 &   1158 &  7 &  9.98 & 13.6 & tf   \\
NGC4562       &J123534.8+255100 &   1315 &  7 & 11.12 & 15.1 & tf   \\
IC3571       &J123620.1+260503 &   1224 & 10 & 15.37 &   ---  &       \\
NGC4565       &J123620.8+255916 &   1192 &  3 &  5.50 & 16.2 & sbf  \\
NGC4670       &J124517.1+270732 &   1046 &  9 & 10.40 &   ---  &      \\
KUG1247+260  &J124940.6+254621 &   1262 &  8 & 14.10 &   ---  &       \\
NGC4725       &J125026.6+253003 &   1180 &  2 &  6.06 & 12.4 & cep  \\
KUG1249+263  &J125144.4+260638 &   1209 & 10 & 13.83 &  9.2 & tf    \\
NGC4747       &J125146.0+254638 &   1162 &  7 & 10.28 & 10.8 & tf   \\
MRK1338      &J125310.1+251642 &   1075 &  9 & 12.78 &   ---  &       \\
\hline
\end{tabular}
\end{table}

\begin{table}
\begin{tabular}{lcrrrrl}\\ \hline
    (1)    &           (2)   &          (3)&    (4)&   (5) &  (6)&   (7) \\
\hline
\multicolumn{7}{c}{Group N4631/N4278}\\
\hline
IC2992       &J120515.9+305120  &   583 &  9 & 12.43 & 12.7 & tf    \\
MAPS-NGP      &J120634.5+312033  &   542 &  8 & 13.81 &   ---  &      \\
NGC4136       &J120917.7+295539  &   576 &  6 &  9.30 &   ---  &      \\
CGCG158-058  &J121707.4+303836  &   690 &  9 & 12.40 &   ---  &       \\
NGC4278       &J122006.8+291651  &   620 & -5 &  7.21 & 16.1 & sbf  \\
MAPS-NGP      &J122017.5+290608  &   673 &  8 & 15.24 &   ---  &      \\
NGC4286       &J122042.1+292045  &   615 &  7 & 10.81 & 11.6 & tf   \\
2MASX         &J122116.6+290222  &   639 &  1 & 13.80 &   ---  &      \\
NGC4308       &J122156.9+300427  &   591 & -5 & 10.61 &   ---  &      \\
UGC07438      &J122219.5+300342  &   669 &  8 & 13.59 &   ---  &      \\
MAPS-NGP      &J122252.7+334943  &   566 & 10 & 15.65 &   ---  &      \\
UGC07457      &J122309.7+292059  &   632 &  1 & 12.81 &   ---  &      \\
IC3247        &J122313.9+285339  &   539 &  7 & 12.11 & 25.4 & tf   \\
MAPS-NGP      &J122357.4+293547  &   739 &  8 & 14.22 &   ---  &      \\
NGC4393       &J122551.2+273342  &   723 &  7 & 12.00 &   ---  &      \\
NGC4414       &J122627.1+311325  &   703 &  5 &  6.71 & 19.9 & tf   \\
NGC4448       &J122815.4+283713  &   627 &  2 &  7.80 & 23.8 & tf    \\
UGC07673      &J123158.1+294233  &   623 & 10 & 12.92 &   ---  &       \\
NGC4559       &J123557.7+275735  &   787 &  6 &  7.47 &  8.1 & tf    \\
KUG1234+299  &J123714.0+293752  &   740 &  8 & 13.78 &   ---  &       \\
KDG178       &J124010.0+323932  &   774 & 10 & 15.14 &   ---  &       \\
CG1042       &J124147.1+325125  &   696 & 10 & 13.57 &   ---  &       \\
NGC4627       &J124159.7+323425  &   812 & -5 &  8.83 &  9.4 & sbf  \\
NGC4631       &J124208.0+323229  &   614 &  7 &  6.46 &  7.31& rgb  \\
MCG+06-28-022 &J124307.1+322926  &   895 &  8 & 13.21 &   ---  &      \\
NGC4656       &J124357.7+321013  &   643 &  9 & 10.90 &  5.4 & tf   \\
UGC07916      &J124425.1+342312  &   614 & 10 & 14.57 &  9.1 & tf   \\
FGC1497      &J124700.6+323905  &   521 &  8 & 15.37 & 23.4 & tf    \\
\hline
\end{tabular}
\end{table}

\begin{table}
\caption{MK- triplets and pairs with $V_{LG} < 3000$ km/s
		 within $ RA=[11.5^h, 13.0^h], Dec.=[+20^{\circ}, +40^{\circ}]$}
\begin{tabular}{lcrrrrr}\\
\hline
    Gal      &        RA,DEC  &  $V_{LG}$& $T$& $K_s$&  $D$ & $[D]$ \\
\hline
    (1)      &        (2)&      (3)&      (4)&   (5)&   (6)&  (7)  \\
\hline
UGC06570     &J113550.0+352007 &  1580 & 8 &10.72 & 14.9  &tf    \\
UGC06603     &J113802.1+351213 &  1617 & 7 &13.01 & 26.2  &tf    \\
	     &                 &       &   &      &       &      \\
NGC3755      & J113633.4+362437&   1557&  5& 10.59&  32.0 & tf   \\
HS1134+3639 & J113654.7+362316&   1583&  9& 14.88&    ---  &       \\
	     &                 &       &   &      &       &      \\
2MASX        & J113901.3+312916&   2705&  4& 11.88&    ---  &      \\
NGC3786      & J113942.5+315433&   2673&  4&  9.33&    ---  &      \\
NGC3788      & J113944.7+315552&   2639&  3&  9.34&    ---  &      \\
	     &                 &       &   &      &       &      \\
KUG1138+327 & J114107.4+322537&   1704& 10& 14.13&    ---  &       \\
MRK0746     & J114129.9+322059&   1684&  9& 14.17&    ---  &       \\
	     &                 &       &   &      &       &      \\
UGC06684     & J114320.9+312718&   1753&  7& 12.12&  32.9 & tf   \\
UGC06684N1& J114332.7+312728&   1794&  9& 12.79&    ---  &         \\
IC2957      & J114537.0+311758&   1752&  2& 11.46&  30.9 & tf    \\
	     &                 &       &   &      &       &      \\
NGC3930      & J115146.0+380054&    921&  5& 11.18&  16.7 & tf   \\
NGC3941      & J115255.4+365911&    922&  0&  7.31&  12.2 & sbf  \\
UGC06955     & J115829.8+380433&    911&  8& 11.34&  14.9 & tf   \\
	     &                 &       &   &      &       &      \\
KDG083      & J115614.5+311816&    617&  8& 11.31&  13.4 & tf    \\
KUG1157+315 & J120016.2+311330&    593&  8& 12.87&  16.1 & tf      \\
	     &                 &       &   &      &       &      \\
LSBCF573-03 & J120942.6+200252&   2416& 10& 12.54&    ---  &       \\
KUG1209+203 & J121157.7+200140&   2291&  9& 12.47&    ---  &       \\
NGC4158      & J121110.2+201033&   2379&  4&  9.73&    ---  &      \\
	     &                 &       &   &      &       &      \\
KUG1209+243 & J121134.9+240144&   2458& 10& 14.63&    ---  &       \\
NGC4162      & J121152.5+240725&   2512&  4&  9.35&  33.9 & tf   \\
	     &                 &       &   &      &       &      \\
NGC4203      & J121505.1+331150&   1078& $-$1&  7.40&  15.1&sbf  \\
UGC07428     & J122202.5+320543&   1125&  8& 12.83&    -  &        \\
	     &                 &       &   &      &       &      \\
IC0777       & J121923.8+281836&   2495&  4& 11.03&  51.0 & tf   \\
KUG1217+288 & J121956.6+283319&   2455&  9& 14.16&   ---  &        \\
	     &                 &       &   &      &       &      \\
KUG1218+387 & J122054.9+382549&    581&  9& 12.97&    ---  &       \\
KDG105      & J122143.0+375914&    582& 10& 15.07&    ---  &       \\
	     &                 &       &   &      &       &      \\
\end{tabular}
\end{table}

\begin{table}
\begin{tabular}{lcrrrrr}\\
\hline
    (1)      &        (2)&      (3)&      (4)&   (5)&   (6)&  (7)  \\
\hline

UGC07584     & J122802.8+223516&    543&  9& 13.75&   9.2 & tf   \\
LSBCF573-01 & J122805.0+221727&    543& 10& 14.56&   7.5 & tf    \\
	     &                 &       &   &      &       &      \\
NGC4793      & J125440.7+285618&   2474&  4&  8.48&  24.4 & tf   \\
KISSR0148   & J125445.2+285529&   2336& 10& 15.60&    ---  &      \\
	     &                 &       &   &      &       &     \\
NGC4789A     & J125405.2+270859&    355& 10& 11.65&   3.91& rgb \\
LSBCD575-05 & J125540.5+191233&    362& 10& 14.06&   5.5 & tf   \\
NGC4826      & J125643.8+214052&    365&  2&  5.31&   4.44& rgb \\
\hline
\end{tabular}
\end{table}

\begin{table}
\caption{Field galaxies within  $RA=[11.5^h,13.0^h], Dec.=[+20^{\circ},+40^{\circ}],
V_{LG} <3000$ km/s}
\begin{tabular}{lcrrrrr}\\ \hline

 Galaxy      &     J2000        &   $V_{LG}$&    $T$&     $K_s$&     $D_{Mpc}$& $[D]$  \\
\hline
  (1)&             (2)&            (3)&         (4)&      (5)&      (6)&  (7) \\
\hline
UGC06499     &  J113011.3+355208 &   2204 &  8 & 13.66 & 38.5 & tf  \\
MRK0424     &  J113027.7+364414 &   1974 &  9 & 12.82 & 29.3 & tf   \\
KUG1127+385 &  J113033.3+381427 &   1926 &  4 & 12.78 &   ---  &      \\
NGC3712      &  J113109.1+283405 &   1527 &  8 & 11.69 & 28.0 & tf  \\
KUG1128+257 &  J113122.1+253005 &   2796 &  7 & 14.16 & 44.7 & tf   \\
UGC06509     &  J113122.7+230655 &   2827 &  7 & 12.49 & 33.9 & tf  \\
UGC06512      & J113144.6+342000 &   1845 &  8 & 12.65 & 28.4 & tf  \\
UGC06517     &  J113202.4+364153 &   2472 &  4 & 11.10 & 41.1 & tf  \\
UGC06526     &  J113239.3+351942 &   1832 &  6 & 11.03 & 18.3 & tf  \\
UGC06531     &  J113249.0+390505 &   1565 &  6 & 12.45 &   ---  &    \\
KUG1130+337 &  J113330.2+333026 &   2570 &  5 & 11.84 &   ---  &      \\
UGC06545     &  J113343.4+323810 &   2586 &  3 & 10.65 &   ---  &     \\
CGCG186-009 &  J113427.6+331044 &   2468 & $-$2 & 11.05 &   ---  &    \\
UGC06561     &  J113526.7+331810 &   2396 &  7 & 13.03 & 33.1 & tf  \\
WAS24       &  J113602.6+322308 &   2709 &  9 & 14.04 &   ---  &      \\
UGC06599     &  J113742.5+240755 &   1496 &  8 & 14.17 & 38.8 & tf  \\
UGC06610     &  J113844.2+334821 &   1826 &  7 & 13.00 & 34.0 & tf  \\
UGC06637     &  J114024.9+282226 &   1787 &  9 & 11.72 & 34.6 & tf  \\
MRK0426     &  J114049.1+351212 &   1504 &  7 & 12.29 & 21.9 & tf   \\
NGC3813      &  J114118.7+363248 &   1456 &  4 &  8.86 & 23.0 & tf  \\
UGC06658     &  J114202.4+320003 &   1434 &  6 & 12.59 &   ---  &     \\
MRK0429     &  J114626.0+345109 &   1365 &  5 & 11.43 &   ---  &      \\
UGC06782     &  J114857.2+235016 &    452 & 10 & 12.67 & 13.7 & bs  \\
UGC06792     &  J114923.4+394619 &    858 &  7 & 11.95 & 20.8 & tf  \\
UGC06817     &  J115053.0+385249 &    251 & 10 & 10.98 &  2.64& rgb \\
MRK0641     &  J115227.4+345340 &   2196 &  9 & 13.95 &   ---  &      \\
UGC6881      &  J115444.7+200320 &    519 & 10 & 14.50 & 16.4 & tf  \\
KDG82        &  J115539.4+313110 &    560 &  8 & 12.41 & 13.6 & tf  \\
BTS076      &  J115844.1+273506 &    451 & 10 & 14.05 &   ---  &      \\
NGC4032      &  J120032.8+200426 &   1173 &  7 & 10.37 & 10.4 & tf  \\
2MASX     &  J120105.9+285943 &   2910 &  3 & 13.44 &   ---  &        \\
BTS089      &  J120147.0+282138 &    799 & 10 & 14.06 &   ---  &      \\
KUG1201+292 &  J120404.7+285853 &    874 &  9 & 12.67 &   ---  &      \\
NGC4080      &  J120451.8+265933 &    533 &  8 & 10.78 & 15.0 & tf  \\
KUG1202+286 &  J120523.3+282156 &    527 &  8 & 13.15 &   ---  &      \\
DDO109       &  J120723.6+394846 &    894 & 10 & 12.72 & 10.2 & tf  \\
UGC07125     &  J120842.3+364810 &   1074 &  8 & 12.64 & 23.1 & tf  \\
MAPS-NGP &  J120921.2+251203 &    719 &  9 & 15.10 &   ---  &         \\
UGC07143     &  J120946.7+250134 &   2520 &  4 & 11.70 & 35.8 & tf  \\
KUG1209+255 &  J121206.5+251833 &   2542 & 10 & 15.13 &   ---  &      \\
\end{tabular}
\end{table}

\begin{table}
\begin{tabular}{lcrrrrr}\\ \hline
\hline
  (1)&             (2)&            (3)&         (4)&      (5)&      (6)&  (7)\\
\hline

NGC4163      &  J121209.2+361009 &    166 & 10 & 12.85 &  2.96& rgb \\
UGC07236     &  J121349.1+241553 &    887 & 10 & 13.81 &   ---  &     \\
UGC07257     &  J121503.0+355731 &    943 &  8 & 12.25 &  8.8 & tf  \\
NGC4204      &  J121514.4+203931 &    782 &  7 & 11.83 &  8.  & txt \\
UGC07321     &  J121734.0+223223 &    345 &  7 & 10.65 & 17.2 & tf  \\
NGC4275      &  J121952.6+273715 &   2276 &  5 & 10.18 &   ---  &     \\
UGC07427     &  J122155.0+350305 &    725 & 10 & 14.09 &  9.7 & tf  \\
NGC4359      &  J122411.1+313118 &   1231 &  5 & 10.80 & 24.0 & tf  \\
UGC07485     &  J122422.2+210936 &    847 &  9 & 13.25 & ---  &     \\
IC3308       &  J122518.2+264254 &    277 &  8 & 13.00 & 12.8 & tf  \\
KK144        &  J122527.9+282857 &    453 & 10 & 14.02 &   ---  &     \\
NGC4395      &  J122548.9+333248 &    313 &  9 &  9.97 &  4.67& rgb \\
IC3341      &  J122623.4+274444 &    338 &  9 & 11.57 & 14.0 & tf   \\
NGC4455      &  J122844.1+224921 &    588 &  7 & 10.63 &  8.4 & tf  \\
UGC07678     &  J123200.4+394955 &    714 &  9 & 12.47 &  6.9 & tf  \\
UGC07699     &  J123248.0+373718 &    520 &  7 & 11.15 & 14.5 & tf  \\
UGC07697     &  J123251.6+201101 &   2468 &  5 & 12.93 & 38.5 & tf  \\
UGC07698     &  J123254.4+313228 &    322 & 10 & 10.63 &  4.85& rgb \\
NGC4509      &  J123306.8+320530 &    906 &  9 & 12.70 & 10.1 & tf  \\
KUG1230+336 &  J123324.9+332103 &    836 & 10 & 13.57 &  5.3 & tf   \\
UGC07719     &  J123400.6+390110 &    707 &  8 & 12.33 &  9.2 & tf  \\
NGC4534      &  J123405.4+353108 &    810 &  8 & 11.21 &  9.8 & tf  \\
MAPS-NGP &  J123521.0+273347 &   1470 &  9 & 14.59 &   ---  &         \\
UGC7774      &  J123622.5+400019 &    553 &  7 & 12.53 & 22.6 & tf  \\
MAPS-NGP   &  J123649.4+333648 &    522 &  8 & 14.48 & 11.7 & tf    \\
UGCA290     &  J123721.8+384438 &    481 & 10 & 13.33 &  6.70& rgb  \\
UGCA292     &  J123840.1+324601 &    306 & 10 & 13.63 &  3.62& rgb  \\
UGC07816     &  J123856.9+380525 &   2252 &  1 & 13.32 &   ---  &     \\
PGC166140    &  J124131.1+394847 &    669 & 10 & 14.69 &   ---  &     \\
IC3687      &  J124215.1+383007 &    385 & 10 & 11.30 &  4.57& rgb  \\
BTS150      &  J124312.8+352447 &   1819 & 10 & 15.00 &   ---  &      \\
UGCA294     &  J124438.3+282819 &    925 &  9 & 12.46 &  9.9 & tf   \\
IC3740       &  J124530.6+204857 &   2602 &  4 & 12.81 &   ---  &     \\
UGCA298     &  J124655.4+263351 &    774 &  9 & 12.42 &   ---  &      \\
DDO147       &  J124659.8+362835 &    351 & 10 & 12.69 &  5.4 & tf  \\
UGC07990     &  J125027.2+282110 &    495 & 10 & 14.59 & 20.4 & tf  \\
IC3840      &  J125146.1+214407 &    536 & 10 & 14.38 &  5.5 & tf   \\
UGC08011     &  J125221.1+213746 &    716 & 10 & 15.49 & 20.0 & tf  \\
UGC08030     &  J125429.4+261818 &    604 & 10 & 15.61 & 10.9 & tf  \\
BTS160      &  J125526.9+325905 &    906 &  7 & 15.20 &   ---  &      \\
UGCA309     &  J125617.8+343917 &    747 & 10 & 13.11 &   ---  &      \\
PGC166151    &  J125901.0+352852 &    723 &  7 & 14.77 &   ---  &     \\
NGC4861      &   J125902.3+345134 &    862 & 10 & 11.76  & 5.7  &tf \\
\hline
\end{tabular}
\end{table}

\begin{table}
\caption{The mean properties of the MK- groups}
\begin{tabular}{lrrrrrrrrrr} \\ \hline
Group  &$N_v$  &$V_{LG}$ & $\sigma_v$  & $R_h$  &$\lg L_K$ &$\lg M_{vir}$& $\lg M/L$ &  $N_D$ &  $DM$  & $\sigma_{DM}$  \\
\hline                                                                        \\
N3900  & 4 & 1745 &   30  & 227  &10.74 & 11.31&  0.57 &   2 & 32.46&  0.10    \\
       &   &      &       & 291  &10.96 & 11.42&  0.46 &     &      &          \\
       &   &      &       &      &      &      &       &     &      &          \\
N4062  & 4 &  736 &   33  & 175  &10.08 & 11.92&  1.84 &   2 & 30.80&  0.02    \\
       &   &      &       & 247  &10.38 & 12.07&  1.69 &     &      &          \\
       &   &      &       &      &      &      &       &     &      &         \\
N4150  & 4 &  211 &   56  &  56  & 8.68 & 11.61&  2.93 &   4 & 31.05&  0.21    \\
       &   &      &       & 310  &10.17 & 12.35&  2.18 &     &      &          \\
       &   &      &       &      &      &      &       &     &      &          \\
N4151  &14 & 1031 &   69  & 348  &11.03 & 12.56&  1.53 &   6 & 30.83&  0.21    \\
       &   &      &       & 363  &11.07 & 12.58&  1.51 &     &      &          \\
       &   &      &       &      &      &      &       &     &      &          \\
N4244  & 8 &  291 &   38  &  76  & 9.71 & 11.60&  1.89 &   8 & 27.95&  0.51    \\
       &   &      &       &  73  & 9.67 & 11.58&  1.91 &     &      &         \\
       &   &      &       &      &      &      &       &     &      &          \\
N4274  &14 &  990 &  102  & 256  &11.22 & 12.70&  1.48 &   9 & 31.06&  0.29    \\
       &   &      &       & 303  &11.36 & 12.77&  1.41 &     &      &          \\
       &   &      &       &      &      &      &       &     &      &          \\
N4565  &11 & 1191 &   83  & 301  &11.83 & 12.98&  1.15 &   6 & 30.73&  0.36    \\
       &   &      &       & 255  &11.69 & 12.91&  1.22 &     &      &          \\
       &   &      &       &      &      &      &       &     &      &          \\
N4631  &28 &  635 &   90  & 243  &11.12 & 12.98&  1.86 &  11 & 30.41&  1.31   \\
       &   &      &       & 333  &11.40 & 13.12&  1.72 &     &      &          \\
\hline
\end{tabular}
\end{table}

\end{document}